\documentclass[aps,twocolumn,superscriptaddress]{revtex4}
\usepackage{amsmath}
\usepackage{amsfonts}
\usepackage{mathrsfs}
\usepackage{graphicx}
\usepackage{times}
\usepackage{color}

\def\be{\begin{equation}}
\def\ee{\end{equation}}
\def\bea{\begin{eqnarray}}
\def\eea{\end{eqnarray}}

\begin{document}

\title{ Anharmonicity Induced Supersolidity In Spin-Orbit Coupled Bose-Einstein Condensates}% Force line breaks with \\

\author{Huan Wang}
\affiliation{Department of Applied Physics, School of Science,
Xi'an  Jiaotong University, Xi'an  710049, Shaanxi, China}
\affiliation{Shaanxi Province Key Laboratory of Quantum Information and Quantum Optoelectronic Devices, Xi'an Jiaotong University, Xi'an 710049, Shaanxi, China}

\author{Shuai Li}
\affiliation{Department of Applied Physics, School of Science,
Xi'an  Jiaotong University, Xi'an  710049, Shaanxi, China}
\affiliation{Shaanxi Province Key Laboratory of Quantum Information and Quantum Optoelectronic Devices, Xi'an Jiaotong University, Xi'an 710049, Shaanxi, China}

\author{Xiaoling Cui}
\affiliation{Beijing National Laboratory for Condensed Matter Physics, Institute of Physics, Chinese Academy of Sciences, Beijing, 100190, China}
\affiliation{Songshan Lake Materials Laboratory , Dongguan, Guangdong 523808, China}

\author{Bo Liu}
\email{liubophy@gmail.com}
\affiliation{Department of Applied Physics, School of Science,
Xi'an  Jiaotong University, Xi'an  710049, Shaanxi, China}
\affiliation{Shaanxi Province Key Laboratory of Quantum Information and Quantum Optoelectronic Devices, Xi'an Jiaotong University, Xi'an 710049, Shaanxi, China}

\begin{abstract}
Supersolid, a fascinating quantum state of matter, features novel phenomena such as the non-classical rotational inertia and transport anomalies. It is a long standing issue of the coexistence of superfluidity and broken translational symmetry
in condensed matter physics. By recent experimental advances to create tunable synthetic spin-orbit
coupling in ultracold gases, such highly controllable atomic systems would provide new possibilities to access supersolidity with no counterpart in solids. Here we report that the combination of anharmonicity of trapping potential and spin-orbit coupling will provide a new paradigm to achieve supersolids. By means of imaginary time evolution of the Gross-Pitaevskii equation, we demonstrate that a supersolid state can be found when considering a trapped Rashba-type spin-orbit coupled bosonic atoms loaded in a one-dimensional optical lattice. Furthermore, a skyrmion-anti-skyrmion lattice is  associated with the appearance of such supersoildity, indicating the topological nontrivial properties of our proposed supersolids.

\end{abstract}

\maketitle

Spin-orbit coupling (SOC) plays a key role in a variety of exotic quantum
phenomena, such as Hall effect and topological insulators~\cite{Xiao_RevModPhys_2010,Nagaosa_RevModPhys_2010,Kane_RevModPhys_2010,zhang_RevModPhys_2011}.
It has been shown that in solid state materials SOC arises from the relativistic
effect when electrons move in a crystal potential. The spin-orbit coupled material is
not only a promising platform to access various fantastic topological states,
but also stimulates a new trend in developing quantum technologies, like quantum computation
and spintronic devices~\cite{Sarma_PhysRevLett_2010,Sarma_RevModPhys_2004}. On the other hand, recent experimental realization of artificial SOC in ultracold atoms supplies new access to control atomic gases and thus opens up a new thrust towards discovering novel quantum states of matter, in particular for that have no prior analogs in traditional condensed matter systems~\cite{2013_Galitski_nature,2011_RevModPhys,2011_Lin_nature,2016_Wu_science,2012_Cheuk_PhysRevLett,2012_Wang_PhysRevLett,
2014_Jotzu_Nature,2014_Bloch_science,2013_Aidelsburger_PhysRevLett,2013_Miyake_PhysRevLett,2013_cheng_Natphys}. One of such desirable unconventional quantum states is the supersolidity, which is characterized by two independent spontaneously broken symmetries, i.e., $U(1)$ and translational symmetries with corresponding superfluidity and density order~\cite{1957_Gross_PhysRev,1958_GROSS}. Such a long-sought quantum state of matter shows a variety of novel properties, such as the non-classical rotational inertia and other anomalous transport features~\cite{1970_legget_PhysRevLett,2006_Clark_PhysRevLett,2008_Ray_PhysRevLett,2009_Boninsegni_PhysRevB}. It thus has attracted tremendous interests in both theoretical and experimental studies in solids and atomic matter systems. For example, a supersolid was predicted to exist in bulk helium, but to prove its existence is still an open question in
recent experiments~\cite{2010_Balibar_nature,2004_Kim_nature,2004_Leggett_science,2005_Nikolay_PhysRevLett,2012_Boninsegni_RevModPhys,2006_Sasaki,2006_Kim_PhysRevLett,2007_Pollet_PhysRevLett,2007_Nikolay}. Thanks to the high controllability in ultracold atomic gases, there have been great interests in
searching such supersolids via ultracold atoms in both experimental and theoretical studies~\cite{2008_Bloch_RevModPhys,2008_Schneider_science,2008_Esslinger_nature,2011_Hemmerich_NatPhys,
2015_Hemmerich_PhysRevLett,2011_RevModPhys,2013_Galitski_nature,2014_Bo_arxiv,2016_Bo_PhysRevA,2016_Bo_PhysRevAI,PhysRevLett_2018_Bo,
2015_Hulet_nature,2017_Greiner_nature,2011_Greiner_nature,2010_Esslinger_Review}. It was previously
predicted to appear in polar molecules, magnetic and Rydberg atoms~\cite{2010_Henkel_PhysRevLett,2010_Cinti_PhysRevLett,2012_Henkel_PhysRevLett,2005_Wessel_PhysRevLett,2009_Danshita_PhysRevLett,
2011_Tieleman_PhysRevA}. In particular, the recent observation of a stripe phase with supersolid properties in spin-orbit-coupled Bose-Einstein Condensates links the SOC with supersolidity~\cite{2017_nature1,2017_nature2}. It indicates that the artificial SOC introduces another degrees of freedom to manipulate atomic gases and provides new opportunities for searching such a novel quantum state.

Here we report the discovery of a new mechanism to achieve the supersolidity by the 'pin effect' resulting from the
anharmonicity of trapping potential. This idea is motivated by the recent discovery of important consequences arising from
the harmonic trap, which is necessary in experiments, when studying the spin-orbit coupled bosons~\cite{Hui_PhysRevLett_2012,Santos_PhysRevLett_2011}. However, so far the effect of anharmonicity of trapping potential has rarely been studied. This work is devoted to unveil the miraculous effect of
anharmonicity. We shall illustrate this with a trapped quasi-two-dimensional spin-$1/2$ interacting Bose gas in the presence of both a Rashba-type SOC and a one-dimensional optical lattice. Such a one-dimensional optical lattice on top of an isotropic $2$D harmonic trap can be considered as an array of anharmonic trapping potentials located at each lattice
depth minimum separately. It turns out that the interplay between anharmonic
traps, SOC and interactions would lead to a novel supersolid. Through imaginary time evolution of the Gross-Pitaevskii equation, we find that for the weak SOC and shallow lattice potential, as expected, a superfluid stripe phase will be
energetically favored~\cite{Zhai_PhysRevLett_2010}. While increasing the SOC and lattice depth, there is a first-order phase transition and a supersolid
with rectangle density profile appears. Furthermore, such a supersolid phase also possesses an exotic topological spin texture, where a skyrmion-anti-skyrmion lattice is formed associated with the appearance of supersoildity.

\textit{Anharmonicity induced 'pin effect' $\raisebox{0.01mm}{---}$} Let us consider a trapped quasi-two-dimensional spin-$1/2$ interacting Bose gas in the presence of a Rashba-type SOC. It can be described by the following model Hamiltonian $\hat H=\hat H_{0}+\hat H_{int}$, where
\begin{equation}
\hat{H}_{0}=\int d^2\mathbf{r}\Psi^{\dag }\left[ \frac{\mathbf{k}^{2}}{2m}+V_{trap}(%
\mathbf{r})+\frac{\kappa }{m}\mathbf{k\cdot \hat{\sigma} }\right] \Psi ,
\label{1}
\end{equation}
is the single-particle Hamiltonian with $\Psi =(\Psi _{\uparrow },\Psi _{\downarrow })^{T}$ denoting the Bose field operators for two pseudospin bosons. The Rashba-type SOC is captured by the last term in $\hat{H}_{0}$ with $\kappa$ describing the strength of SOC and $\hat{\sigma}$ being the Pauli matrix. The interaction part of $\hat H$ can be expressed as
\begin{equation}
\hat{H}_{int}=\int d^2\mathbf{r}\left( g_{1}\hat{n}_{\uparrow }^{2}+g_{2}\hat{n%
}_{\downarrow }^{2}+2g_{12}\hat{n}_{\uparrow }\hat{n}_{\downarrow
}\right) ,  \label{2}
\end{equation}
where the density operators for two pseudospin bosons are defined as $\hat{n}_{\uparrow }=\Psi _{\uparrow }^{\dag }\Psi _{\uparrow },\hat{n}_{\downarrow }=\Psi _{\downarrow }^{\dag }\Psi _{\downarrow }$ respectively.  $g_{1}$ and $g_{2}$
characterize the intraspecies contact interaction strengths, while $g_{12}$ labels the interspecies contact interaction strength. Those are determined by the effective intraspecies and interspecies $s$-wave scattering length respectively.
In this work, we shall focus on the case with $g_1 = g_2 > 0$. When further considering the presence of an isotropic harmonic trap $V_{trap}(\mathbf{r})=\frac{1}{2}m\omega^{2}(x^{2}+y^{2})$ mimicking the practical situation in cold atom experiments, it is known that the ground state is a superfluid stripe phase in the region where $c_2/c_0<0$ with $c_0 = g_1 + g_{12}$ and $c_2 = g_1-g_{12}$~\cite{Zhai_PhysRevLett_2010}. Due to the spatially isotropic property of the harmonic trap, the stripe phase along different directions are energetically degenerate, where the direction of stripe phase is determined by the condensate momentum. To break this spatially rotational symmetry, one can consider utilizing the anisotropy of trapping potential. Interestingly, we find that the anharmonic trap will pin the direction of density stripe.

\begin{figure}[t]
\includegraphics[width=8cm]{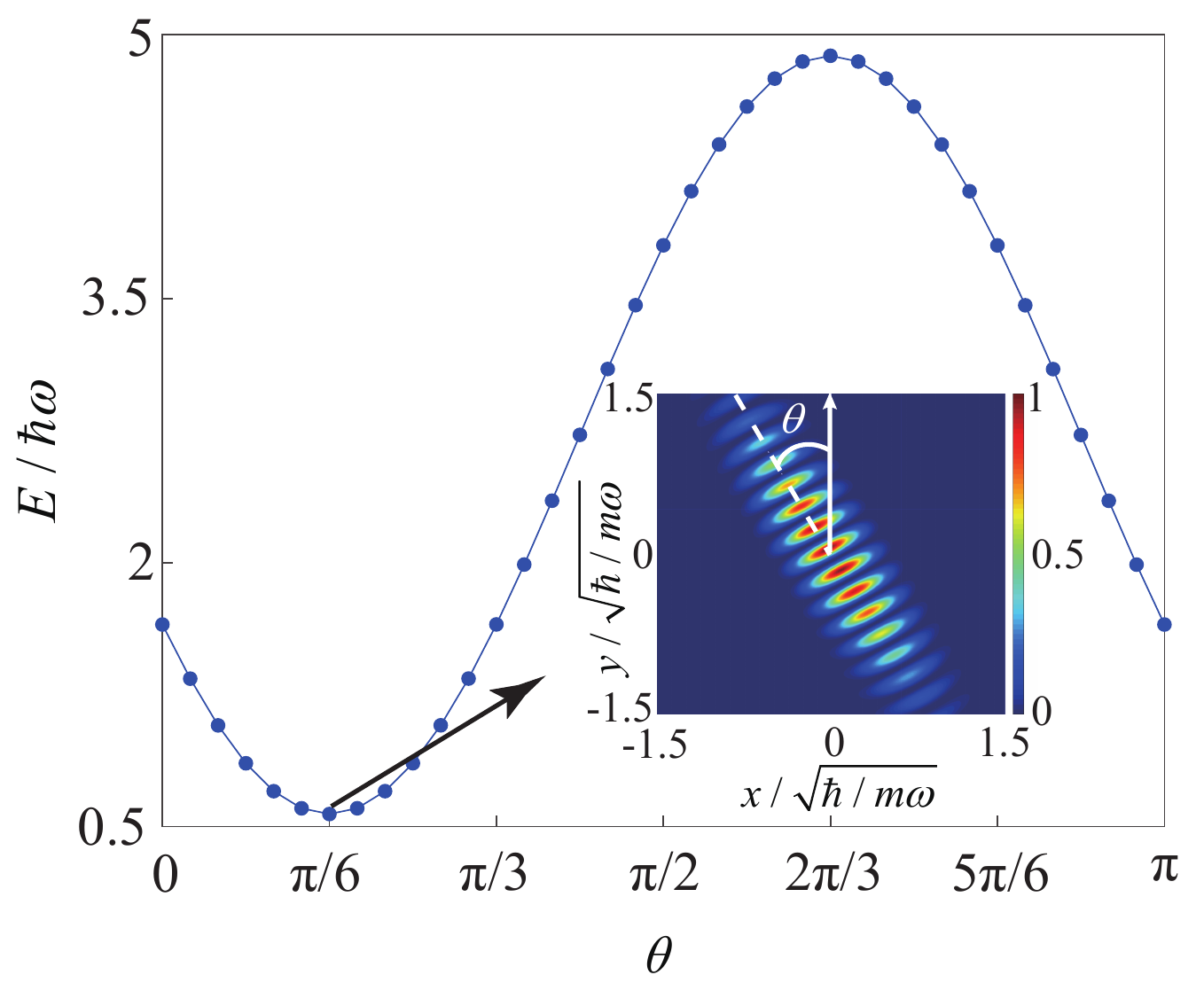}
\caption{The expectation value of the energy E contributed from the anharmonic trap $V_{Atrap, \theta_{0}}(\mathbf{r})$
for different superfluid stripe phases pointing along the direction located at the angle $\theta$ with respect to the vertical direction. Here we choose the long axis of the anharmonic trap located at the angle $\theta_{0}=\pi/6$.
It is shown that the stripe phase pointing along the long axis of the trapping potential mostly minimizes the energy cost compared to all the other direction-pointing stripe phases. {The inset shows the stripe phase pointing along
$\theta=\pi/6$, where the dashed-line indicates the direction determined by the polar angle of the condensate
momentum. The polar axis here is set as the positive $y$-axis.} Other parameters are chosen as $\gamma=15$, $\kappa'=14$, $c'_{0}=2$ and $c'_{2}=-0.8c'_{0}$.}
\label{fig:Pin}
\end{figure}

To demonstrate this, let us consider compressing the isotropic harmonic trap into a cigar shape and the corresponding trapping potential can be written as
$V_{Atrap}(\mathbf{r})=\frac{1}{2}m\omega^{2}(\gamma x^{2}+y^{2})$ with $\gamma$ describing the anisotropy. When $\gamma>1$, the trapping potential becomes elliptical with the long axis residing in the $y$-direction (vertical direction) and short axis along the $x$-direction. Without loss of generality, we further set the long axis of the elliptical trapping potential located at a certain angle $\theta_{0}$ with respect to $y$-axis. Then the ground state can be found numerically by minimizing the following dimensionless energy functional constructed under the Gross-Pitaevskii mean-field theory
\begin{eqnarray}
\varepsilon &=& \int d^{2} \mathbf{r} \sum_{\sigma=\uparrow, \downarrow} \psi_{\sigma}^{*}\left[-\frac{1}{2}\nabla^{2}+ V_{Atrap, \theta_{0}}(\mathbf{r}) \right] \psi_{\sigma} \notag \\
&+&\kappa'\left[\psi_{\uparrow}^{*}\left(-i \partial_{x}-\partial_{y}\right) \psi_{\downarrow}+\psi_{\downarrow}^{*}\left(-i \partial_{x}+\partial_{y}\right) \psi_{\uparrow}\right] \notag \\
&+&\frac{c'_{0}}{2}\left(\left|\psi_{\uparrow}\right|^{2}+\left|\psi_{\downarrow}\right|^{2}\right)^{2}+\frac{c'_{2}}{2}\left(\left|\psi_{\downarrow}\right|^{2}-\left|\psi_{\uparrow}\right|^{2}\right)^{2} . \label{3}
\end{eqnarray}
Here we choose $\hbar\omega$, $\sqrt{\hbar/m\omega}$ and $1/\omega$ as the units of energy, spatial length, and time respectively.
$V_{Atrap, \theta_{0}}(\mathbf{r})=\frac{1}{2}(\gamma
{(\cos\theta_{0}x+\sin\theta_{0}y)}^{2}+{(\cos\theta_{0}y-\sin\theta_{0}x)}^{2})$ represents the anharmonic trap with
the long axis located at the angle $\theta_{0}$ with respect to the vertical direction. The dimensionless interaction strength is defined as $c'_{0}=\beta_{1}+\beta_{12}$ and $c'_{2}=\beta_{1}-\beta_{12}$, where $\beta_{1}=g_{1}N m/{\hbar}^2$ and $\beta_{12}=g_{12}N m/{\hbar}^2$ with the total particle number $N$. And the dimensionless SOC strength is
$\kappa'=\kappa/\sqrt{\hbar m\omega}$. What we found is shown in Fig.~\ref{fig:Pin}. For example, in the presence of an anharmonic trap with the long axis located at the angle $\theta_{0}=\pi/6$, the stripe phase pointing the long axis of the trap mostly minimizes the energy resulting from the trapping potential compared to all the other direction-pointing stripe phases. Since the kinetic energy and interaction energy are the same for the stripe phases pointing along different directions, the anharmonic trap will pin the direction of the stripe phase along the long axis of the trap, where the system will be energetically favored.

\begin{figure*}[t]
\includegraphics[width=18cm]{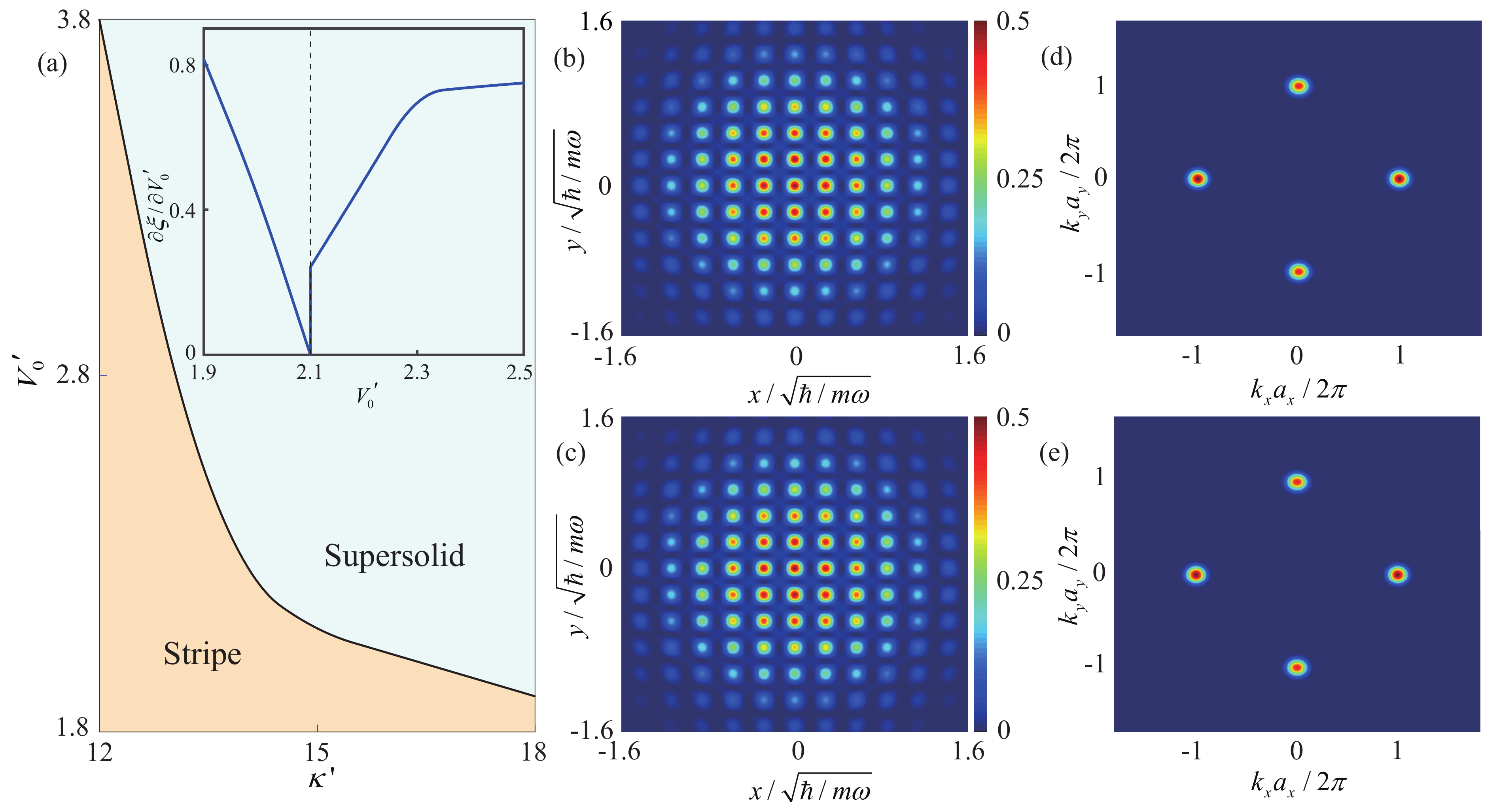}
\caption{(a) Zero-temperature phase diagram of the model Hamiltonian in Eq.~\eqref{4} as a function of the SOC strength $\kappa'$ and lattice  depth $V'_{0}$. For certain SOC, there is a threshold of the lattice depth, beyond that the supersolid phase appears. The inset shows the nonanalytic behavior of the first-order derivative of the ground-state energy density $\xi$. Panels (b) and (c) show the density profile of the supersolid phase for spin-up and down components, respectively. It is shown that in the supersolid phase a spatially periodic rectangle density distribution is formed. And the periodicity of such a density profile is determined by the period of lattice potential and SOC for the horizontal and vertical directions, respectively. (d) and (e) show the corresponding momentum density distribution with respect to (b) and (c). To demonstrate the periodic structure of the supersolid, in (d) and (e) we just show the oscillatory
portion of the density distribution by eliminating the effect of the background average density profile. Other parameters are chosen as $c'_{0}=10$, $c'_{2}=-0.8c'_{0}$, $k^{\prime}_L=3.5\pi$, $\kappa'=16$, $V'_{0}=8$.} \label{fig:Zphase}
\end{figure*}

\begin{figure}[t]
\includegraphics[width=9.2cm]{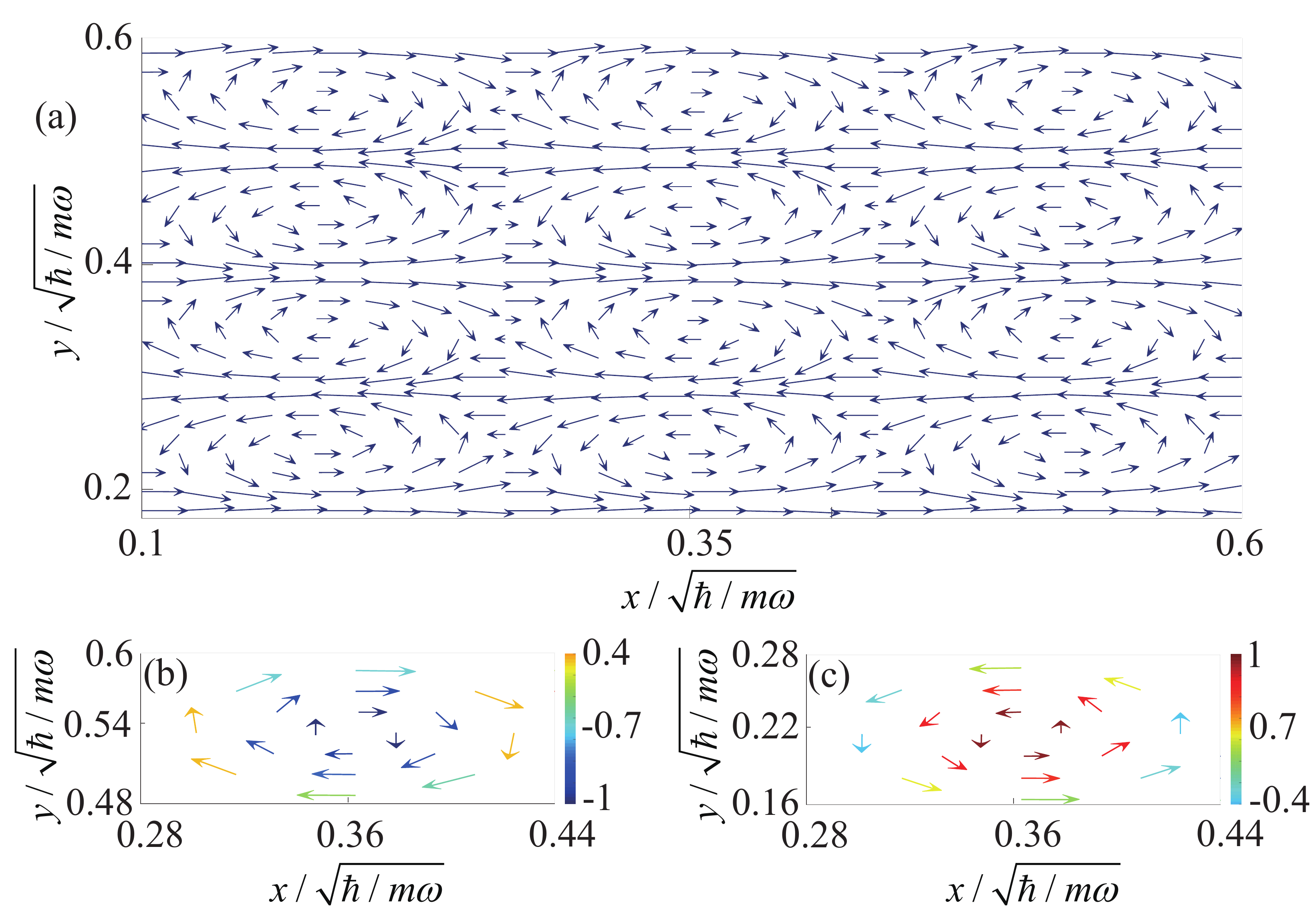}
\caption{The skyrmion-anti-skyrmion lattice spin texture of the proposed supersolid phase. (a) Illustration of the skyrmion-anti-skyrmion lattice configuration formed by the vector $S$ defined in the main text.
It is shown that the $x$ and $y$ components of $S$ forms a vortex-anti-vortex lattice structure in $xy$-plane.
(b) and (c) show the skyrmion and anti-skyrmion respectively. Here the arrows show ($S_{x},S_{y}$)
and the colors index the z component of $S$. Other parameters are chosen as $c'_{0}=20$, $c'_{2}=-0.8c'_{0}$, $k^{\prime}_L=8\pi$, $\kappa'=30$, $V'_{0}=7$.} \label{fig:Spin}
\end{figure}

\textit{Anharmonicity induced supersolidity $\raisebox{0.01mm}{---}$} In the previous section, we demonstrate that the anisotropy of a single trapping potential will break the spatially rotational symmetry and pin the direction of the stripe phase. By utilizing such an anharmonic 'pin effect', an unexpected scheme through
constructing an array of anharmonic trapping potentials to achieve the supersolidity is surprisingly
unveiled. To be more specific, a concrete model will be introduced as follows. Let us still consider a trapped quasi-two-dimensional spin-$1/2$ interacting Bose gas in the presence of a Rashba-type SOC. However, distinguished from the case in the previous section, the new ingredient is to add a one-dimensional optical lattice. Therefore, such a system can be described by the following model Hamiltonian
\begin{eqnarray}
{\hat {H}} &=& \int d^2\mathbf{r}\Psi^{\dag }\left[ \frac{\mathbf{k}^{2}}{2m}+V_{trap}(\mathbf{r})+V_{OL}(\mathbf{r})+\frac{\kappa }{m}\mathbf{k\cdot \hat{\sigma} }\right] \Psi   \notag \\
&+& \int d^2\mathbf{r}\left( g_{1}\hat{n}_{\uparrow }^{2}+g_{2}\hat{n}_{\downarrow }^{2}+2g_{12}\hat{n}_{\uparrow }\hat{n}_{\downarrow}\right),
\label{4}
\end{eqnarray}
where $V_{OL}(\mathbf{r})=V_{0}\sin ^{2}(k_{L}x)$ is a $1$D optical lattice along the $x$-direction.
$V_{0}$ is the lattice depth and $k_{L}$ is the wavevector of the laser field with the corresponding lattice constant defined as $a_x=\pi/k_{L}$. Other parameters defined in Eq.~\eqref{4} are the same as in Eq.~\eqref{1} and ~\eqref{2}. By choosing the same units as in Eq.~\eqref{3}, we obtain the following dimensionless energy functional under the Gross-Pitaevskii mean-field theory
\begin{eqnarray}
\varepsilon &=&\int d^{2} \mathbf{r} \sum_{\sigma=\uparrow, \downarrow} \psi_{\sigma}^{*}\bigg[-\frac{1}{2} \nabla^{2}+\frac{1}{2}\left(x^{2}+y^{2}\right) \notag \\
&+&V'_{0}\sin ^{2}\left(k_{L}^{\prime} x\right)\bigg] \psi_{\sigma}+\kappa' \big[\psi_{\uparrow}^{*}\left(-i \partial_{x}-\partial_{y}\right) \psi_{\downarrow} \notag \\
&+&\psi_{\downarrow}^{*}\left(-i \partial_{x}+\partial_{y}\right) \psi_{\uparrow} \big] +\frac{c'_{0}}{2}\left(\left|\psi_{\uparrow}\right|^{2}+\left|\psi_{\downarrow}\right|^{2}\right)^{2} \notag \\
&+&\frac{c'_{2}}{2}\left(\left|\psi_{\downarrow}\right|^{2}-\left|\psi_{\uparrow}\right|^{2}\right)^{2} ,
\label{5}
\end{eqnarray}
where $V'_{0}=V_{0}/\hbar\omega$ is the dimensionless lattice depth and the dimensionless wavevector is defined as
$k_{L}^{\prime}=k_{L}\sqrt{\hbar/m\omega}$. The defined $c'_{0}$ and $c'_{2}$ are the same as in Eq.~\eqref{3}.

Through numerically computing the ground state via minimizing the dimensionless energy functional in Eq.~\eqref{5} by using imaginary time evolution method, the phase diagram is obtained as shown in Fig.~\ref{fig:Zphase}(a). {Note that
the phase diagram here is constructed in the weakly interacting regime, where the kinetic energy is much larger than both intraspecies and interspecies interaction energy. It is confirmed in our numerics that the interaction energy is typically smaller than the kinetic energy by two orders of magnitudes. Therefore, in such a region the Gross-Pitaevskii mean-field theory is valid.} Here we also focus on the case with the interaction $c_2/c_0<0$. As shown in Fig.~\ref{fig:Zphase}(a), there are two different phases in the phase diagram, which consists of a superfluid stripe phase and a supersolid phase. A threshold of lattice depth separates the above two different phases when considering a fixed
Rashba SOC strength. Below that lattice depth threshold, the ground state of the system is a superfluid stripe phase
where the translational symmetry along the $y$-direction is spontaneously broken. While further increasing the lattice
depth above the critical value, the translational symmetry along the $x$-direction is also broken resulting from the periodic lattice potential along the $x$-direction. As a result, a new superfluid with spontaneously formed periodic density modulations along both $x$ and $y$ directions is obtained as shown in Fig.~\ref{fig:Zphase} (b) and (c) for the two pseudospin bosons respectively, where the translational symmetry in $2$D-plane and $U(1)$ gauge symmetry are simultaneously broken. Therefore, it can be considered as a supersolid state. {The transition between the above two different phases is a first-order phase transition, which is identified by the nonanalytic behavior of the first-order derivative of the ground-state energy density as shown in the inset of Fig.~\ref{fig:Zphase}(a). It is also shown that the threshold of lattice depth monotonically decreases when increasing the strength of SOC, which can be understood from the fact that
the kinetic energy cost resulting from the density modulation induced by adding the optical lattice for the
case with weaker SOC is much larger than that for the one with stronger SOC. For example, we numerically calculate the kinetic energy cost when enlarging the lattice depth, starting from the same lattice depth $V'_0=1$ to the lattice depth thresholds for different SOC strengths $\kappa'=11$ and  $\kappa'=17$, respectively. And we find that the kinetic energy cost
for $\kappa'=11$ is much larger than that for $\kappa'=17$ by two orders of magnitudes. Therefore, to form a supersolid, the weaker the SOC is, the deeper the optical lattice will be needed to reduce the energy cost.}

Here we would like to stress the understanding through the anharmonic 'pin effect' introduced above to explain the appearance of our proposed supersolids. In our scheme, the presence of the one-dimensional optical lattice on top of an isotropic $2$D harmonic trap can be considered as an array of anharmonic trapping potentials elongated along the vertical direction, which are correspondingly located at each lattice depth minimum. Therefore, the stripe phase will be pinned along the vertical direction and the translational symmetry in the $y$-direction is thus spontaneously broken. When further increasing the lattice depth above the threshold as shown in Fig.~\ref{fig:Zphase}(a), a periodic density modulation coincided with the period of lattice potential along the horizontal direction is formed and thus the translational symmetry along the $x$-direction is also broken. Such a superfluid with both the translational symmetry in $2$D-plane and $U(1)$ gauge symmetry simultaneously breaking can be considered as a supersolid. This is very different from the usual manner of creating supersolid by rotation or artificial gauge fields~\cite{Henkel_PhysRevLett_2012,Tieleman_PhysRevA_2011}.

Furthermore, a characteristic feature of the momentum density distribution can be used to distinguish a supersolid from a superfluid stripe phase, which can be detected using conventional time-of-flight imaging technique. Specifically, as shown in
Fig.~\ref{fig:Zphase} (d) and (e), in the supersolid phase the peaks of
momentum density distribution are located at both $(\pm\frac{2n\pi}{a_x},0)$ and  $(0,\pm\frac{2m\pi}{a_y})$, where
$m, n$ are non-zero positive integers. $a_x$ and $a_y$ are the periodicity of the spatial density distribution along the $x$ and $y$ directions respectively, which correspondingly depend on the periodicity of the lattice potential and SOC strength. Such a feature of the momentum density distribution in supersolids is dramatically distinguished from that of the
superfluid stripe phase, where the momentum density distribution peaks are only located along the $y$ direction
at $(0,\pm\frac{2n\pi}{a_y})$ for the case we considered here.

Besides the above characteristic feature of the momentum density distribution, the supersolid phase proposed here also exhibits exotic spin textures, which will be discussed below. To demonstrate that, let us firstly define a spin density vector for the spin-$1/2$ bosons as $\mathbf{S}=\mathbf{\Psi}^{\dag}\boldsymbol{\sigma}\mathbf{\Psi}/|\mathbf{\Psi}|^{2}$,
with $\boldsymbol{\sigma}$ representing the Pauli matrix. As shown in Fig.~\ref{fig:Spin}(a),
the spin texture represents a periodic magnetic structure accompanying with the emergence of the supersolid phase.
If we zoom in on such a spin texture, there are two different structures. As shown in Fig.~\ref{fig:Spin} (b) and (c), the
vector $\mathbf{S}$ wraps around a sphere and it points to the south (north) pole in the center, while with increasing radius, it varies continuously and eventually points to the north (south) pole. Therefore, the vector $\mathbf{S}$ forms Skyrmions (anti-Skyrmions). It is also shown that the Skyrmion and anti-Skyrmion spin textures appear at where the spin-down and up components are centralized respectively. And the cores of the spin structure as shown in
Fig.~\ref{fig:Spin}(a) are correspondingly centered around the minimum of density distributions for spin-up and down components along the
horizontal direction. Such spin textures are thus formed a skyrmion-anti-Skyrmion lattice coincided with the appearance of
the supersolid. Our scheme hence provides a new way to create and manipulate exotic spin textures in spin-orbit coupled ultracold gases.

We next demonstrate that the proposed supersolid here indeed features topologically nontrivial spin textures.
The topological nature of the above spin structures can be characterized by the topological charge $Q$,
which can be defined as a spatial integral of the topological charge density
\begin{equation}
Q=\int d^{2} \mathbf{r} \frac{1}{4 \pi} \mathbf{S} \cdot\left(\frac{\partial \mathbf{S}}{\partial x} \times \frac{\partial \mathbf{S}}{\partial y}\right). \label{8}
\end{equation}
Through numerics we find that a Skyrmion carries a topological charge $1$, while an anti-Skyrmion carries a topological charge $-1$, which is distinguished from the topologically trivial case where the topological charge is zero.

\textit{Conclusion $\raisebox{0.01mm}{---}$} In summary, we have demonstrated a new approach to achieve the supersolidity via the combination of SOC and anharmonicity of trapping potential in a trapped spin-orbit coupled ultracold bosons loaded in a one-dimensional optical lattice. The crucial ingredient of our scheme is to engineer the anharmonicity of trapping potential. The long-sought supersolid phase has been unveiled, which also shows an exotic spin texture, i.e., a skyrmion-anti-skyrmion spin lattice. Our approach is rather generic to the spin-orbit coupled quantum gases than restricted to the setup considered in this work. It thus complements with a new window in cold gases to realize and furthermore to control the various supersolids.

\textit{Acknowledgment $\raisebox{0.01mm}{---}$} This work is supported by the National Key Research and Development Program of China (2018YFA0307600, 2016YFA0300603) and NSFC Grants No. 11774282, 11622436, 11421092, 11534014.

\bibliographystyle{apsrev}
\bibliography{supersolid}

\end{document}